\documentstyle[12pt]{article}
\newcommand{\be}{\begin{equation}
\setlength{\abovedisplayskip}{4.5mm}
\setlength{\belowdisplayskip}{4.5mm}
\setlength{\abovedisplayshortskip}{3mm}
\setlength{\belowdisplayshortskip}{3mm}}
\newcommand{\ber}{\begin{eqnarray}
\setlength{\abovedisplayskip}{4.5mm}
\setlength{\belowdisplayskip}{4.5mm}
\setlength{\abovedisplayshortskip}{3.8mm}
\setlength{\belowdisplayshortskip}{3.8mm}}
\newcommand{\ba}{\begin{eqnarray}
\setlength{\abovedisplayskip}{4.5mm}
\setlength{\belowdisplayskip}{4.5mm}
\setlength{\abovedisplayshortskip}{3.8mm}
\setlength{\belowdisplayshortskip}{3.8mm}}
\newcommand{\ea}{\end{eqnarray}}
\newcommand{\ee}{\end{equation}}
\newcommand{\eer}{\end{eqnarray}}

\newcommand{\zbar}{{\overline{z}}}

\newcommand{\taubar}{{\overline{\tau}}}

\newcommand{\ie}{{\it i.e.\ }}
\newcommand{\1}{{\it 1}}
\newcommand{\Z}{{\bf Z}}

\newcommand{\R}{{\bf R}}
\newcommand{\Aa}{{\rm \bf A}}
\newcommand{\Bb}{{\rm \bf B}}

\newcommand{\ra}{\rightarrow}

\newlength{\minusspace}
\newlength{\minuslength}
\renewcommand{\Im}{{\rm Im}}
\renewcommand{\Re}{{\rm Re}}

\newcommand{\newsection}[1]{
\vspace{10mm}

\addtocounter{section}{1}
\setcounter{equation}{0}
\setcounter{subsection}{0}
\begin{flushleft}
{\large\bf \thesection. #1}
\end{flushleft}
\noindent}

\newcommand{\newsubsection}[1]{
\medskip

\addtocounter{subsection}{1}
\begin{flushleft}
{\bf \thesection.\alph{subsection}. #1}
\end{flushleft}
\noindent}
\newcommand{\bbox}[2]{\raise.7ex \hbox{{${\scriptstyle
\;#1}$}\kern-.25em\lower.5ex \hbox {\vtop
 {\hbox to 5mm{\hfil\blok\hfil} \kern-2.2ex \hbox to 5mm{\hfil
{$\scriptstyle #2$}\hfil}}}}}
\newcommand{\blok}{\hbox{\vrule \vbox
{\hrule \kern2.5mm\hbox{\kern2.5mm}\hrule}\vrule}}
\newcommand{\BBox}[2]{\hbox{$\ #1\;\lower.8ex\hbox {\vtop
 {\Blok \kern-1.8ex \hbox to 5mm{\hfil $#2$\hfil}}}\kern.2em$}}
\newcommand{\Blok}{\hbox{\vrule
 \vbox {\hrule \kern4mm\hbox{\kern4mm}\hrule}\vrule}}
\def \Block#1{{\kern.25em\lower.8ex\hbox {\vtop
 {{\hbox{\vrule \vbox {\hrule \kern4mm\hbox{\kern4mm}\hrule}\vrule}}
  \kern-1.2ex \hbox to 5mm{\hfil $#1$\hfil}}}}}

\newcommand{\hf}{{1\over2}}

\begin{document}
\addtolength{\baselineskip}{.5mm}

\newcommand{\opi}{{1\over 2\pi}}
\newcommand{\nonu}{\nonumber\\}
\newcommand{\is}{\!& = &\!}
\newcommand{\r}{\rangle}
\newcommand{\Bl}{\Bigl\langle}
\newcommand{\Br}{\Bigr\rangle}
\newcommand{\del}{\partial}
\newcommand{\XX}{{M^4}}

\newcommand{\Eij}{L(C_i, C_j)}
\newcommand{\Gij}{G(C_i, C_j)}
\newcommand{\Bij}{B(C_i, C_j)}
\newcommand{\Gijt}{G(C_1, C_2)}
\newcommand{\Bijt}{B(C_1, C_2)}
\newcommand{\oversea}{{\textstyle{1\over C}}}

\thispagestyle{empty}

\begin{center}
{\large\sc{Global Aspects of Electric-Magnetic Duality\\[27mm]}}
{\sc Erik Verlinde}\\[5mm]
{\it TH-Division, CERN\\[2mm]
CH-1211 Geneva 23}\\[3.5mm]
and\\[3.5 mm]
{\it Institute for Theoretical Physics\\[2mm]
University of Utrecht\\[2mm]
P.O. BOX 80.006, 3508 TA Utrecht}\\[23mm]

{\sc Abstract}
\end{center}

\vspace{-1mm}
\noindent
We show that the partition function of free Maxwell theory
on a generic Euclidean four-manifold transforms
in a non-trivial way under electric-magnetic duality. The classical
part of the  partition sum  can be mapped onto the genus-one
partition
function of a 2d toroidal model, without the oscillator
contributions.
This map relates electric-magnetic duality to modular invariance of
the
toroidal model and, conversely, the $O(d,d',\Z)$ duality
to the invariance of Maxwell theory under the 4d mapping class group.
These dualities and the relation between toroidal models and
Maxwell theory can be understood by regarding both theories as
dimensional reductions of a self-dual 2-form theory in six
dimensions.
Generalizations to more $U(1)$-gauge fields and reductions from
higher
dimensions are also discussed. We find indications
 that the Abelian gauge theories related to 4d string theories with
$N=4$
space-time supersymmetry are exactly duality invariant.

\vspace{0.5cm}

\begin{flushleft}
{\sc CERN-TH}/95-146\\
May 1995
\end{flushleft}
\vspace{.5cm}

\newpage

\newsection{Introduction}
The purpose of this paper is to study global aspects
of electric-magnetic duality in a very simple model, namely
(source-free) Maxwell theory, and to investigate its relation
with other known duality symmetries  in lower and higher
dimensional free field theories.
Maxwell theory is the simplest example of a 4d
field theory exhibiting a strong-weak coupling duality, and,
just as the toroidal conformal models in two dimensions, it
serves as a useful `toy'-model  to illustrate and
gain more insight in this phenomenon.
Of course, we hope that our results will eventually shed new light on
the recent developments in strong-coupling supersymmetric
non-abelian gauge theories \cite{SeiWit,N=4}, and the
newly discovered dualities in string theory \cite{sdual,
HullTownsend}.
But, just to keep things simple, we will in this paper restrict our
attention to abelian gauge theories, and leave possible
generalizations, applications  or implications to future work
\footnote{We became aware that
some related work has been done \cite{Wittenprep},
while this paper was being proof-read.}.

In the first part of this paper we consider  the partition function
of Maxwell theory on a euclidean four-manifold without boundary,
and exhibit its behaviour as a function of the coupling constants
$g$ and $\theta$ that appear in the euclidean Maxwell-action
\be
\label{action}
S[A]=
{1\over g^2} \int_{\strut \XX}\! F\wedge {}^\ast F-i{\theta\over
8\pi^2}
\int_{\strut \XX}\! F\wedge F.
\ee
Here $A=A_i d x^i$ is the gauge potential written as a one-form,
$F=dA = \hf F_{ij} dx^i\wedge dx^j$ is the 2-form field strength and
${}^\ast \! F= {1\over 4} \sqrt{g}\epsilon_{ij}{}^{kl}
F_{kl}dx^i\wedge dx^j$
denotes its Hodge-dual. We will study the properties of the
partition function under the $SL(2,\Z)$ duality group \cite{sl2}
\be
\label{SL2Z}
\tau \ra {a\tau+b\over c\tau+d}
\ee
where
$
\tau ={\theta\over 2\pi} + {4\pi i\over g^2}
$
and $a,b,c$ and $d$ are integers satisfying $ad-bc=1$.
We will find that only for certain four-manifolds the full
$SL(2,\Z)$-symmetry can be realized. For these manifolds we also
find a curious correspondence with 2d toroidal models, that
interchanges
the role of the duality and mapping class group.  This will be
discussed
and illustrated with examples in section 3. Here we also
show that, in analogy with 2d conformal field theory, the correlators
of the Wilson-'t Hooft line operators can be obtained as a
degenerate limit of the partition function.

The relation between Maxwell theory and the 2d toroidal models is
further
clarified in section 4. Here we show that both theories correspond to
a dimensional reduction of the same 6d
theory describing a $2$-form with self-dual field strength, and that
the
duality symmetries arise from the mapping class group of
the internal manifold. We also find that the partition function
corresponds to a wave-function in a topological theory on which
duality acts as a canonical transformation. Finally,
in section 4.3 we discuss some string related Abelian gauge theories.

\newsection{Duality of the Partition Function}
\vspace{-6mm}
\newsubsection{The classical partition sum}
We consider the partition function of Maxwell theory on a closed
four-manifold
\be
Z={1\over |{\cal G}|}\int [dA] e^{-S[A]}.
\ee
Here we integrate over all $U(1)$ gauge-fields $A$ on the
four-manifold $\XX$
and we divide as usual by the volume of the gauge group $|{\cal G}|$.
Applying the standard Faddeev-Popov procedure to factor out this
volume
one finds that the partition function factorizes into  a sum over the
classical
saddle-points times a product of
determinants. Explicitly,
\be
Z = {\det' \Delta_{FP}\over(\det' \Delta_A)^{1\over 2}} Z_{cl},
\ee
where $Z_{cl}$ represents the contribution of the classical
saddle-points
\be
Z_{cl} =\sum_{saddle\atop points} e^{-S[A_{cl}]}.
\ee
Here $\Delta_{FP}$ and $\Delta_A$ denote the kinetic operators for
the
Faddeev-Popov ghosts and the gauge field and, after gauge fixing,
are given by the Laplacian   acting on functions
or one-forms respectively. Both these laplacians can have
zero modes, which have to be projected out. In particular, when
the four-manifold $M^4$ is non-simply-connected, the Laplacian
$\Delta_A$ has zero-modes corresponding to the flat abelian
connections on $\XX$. In the following we will for simplicity
consider
simply connected manifolds, so that we do not have to deal with
these zero modes.

We will now focus our attention on the sum over classical solutions.
When the four-manifold $\XX$ has non-trivial homology two-cycles, \ie
closed surfaces that do not correspond to the boundary of a
3-dimensional
sub-manifold in $\XX$, there exist
field configurations with non-zero flux through these surfaces.
A generalization of the familiar Dirac quantization condition implies
that the flux through the non-trivial homology two-cycles $\Sigma_I$
must
be quantized
\be
\int_{\Sigma_I} F =2\pi m^I;\qquad m^I\in \Z
\ee
with $I =1,\ldots, \dim H_2(M^4)$.
This tells us that in the absence of sources
the solutions of the field equations
$
d^\ast F= 0
$
can be decomposed as
\be
\label{Fsum}
F= 2\pi \sum_I  m^I \alpha_I
\ee
where $\alpha_I$ is an integral basis of harmonic $2$-forms, which by
definition satisfy $d\alpha_I=d^*\alpha_I=0$ and are normalized so
that
$\int_{\Sigma_I}\alpha_J = \delta^I{}_J$.
Thus on a compact four-manifold the classical saddle-points are
uniquely labeled by the integer magnetic fluxes $m^I$.
Inserting the expression (\ref{Fsum}) into
(\ref{action}) we find that the classical action for this field
configuration is
\be
\label{Sm}
S[m^I] = {4\pi^2\over g^2} m^I G_{IJ} m^J -{i\over 2}\theta m^I
Q_{IJ} m^J
\ee
where
\be
Q_{IJ} =  \int_\XX\! \alpha_I\wedge\alpha_J,  \qquad
G_{IJ}= \int_\XX\! \alpha_I\wedge {}^\ast\alpha_J
\ee
represent the intersection form and the
metric on the space of harmonic two-forms.
In this way we find that the saddle-point contribution to
the partition sum is given by
\be
\label{Zcl}
Z_{cl}(g,\theta) = {\oversea} \sum_{m^I} e^{- S[m^I]},
\ee
where $C$ is a normalization constant.
Thus, while the full partition function $Z$ depends on the detailed
geometry of $\XX$, we find that its classical part
$Z_{cl}$ is completely determined by the matrices $G_{IJ}$ and
$Q_{IJ}$,
and thus requires relatively little information.
For any manifold  $G_{IJ}$ is symmetric and positive-definite, and
the intersection form $Q_{IJ}$ has integer entries and determinant
equal
to one: such matrices are called unimodular.
Its inverse $Q^{IJ}$ counts the number of intersection points
 of the two surfaces $\Sigma_I$ and $\Sigma_J$:
$Q^{IJ}=\#(\Sigma_I,\Sigma_J).$
Further, it follows from ${}^\ast({}^\ast\alpha_I)=\alpha_I$ that
\be
\label{QG}
Q^I{}_K Q^K{}_J = \delta^I{}_J
\ee
where $Q^I{}_J\!\equiv\! G^{IK}Q_{KJ}$. Thus the eigenvalues of
$Q^I{}_J$
are all $+1$ or $-1$ corresponding to the self-dual and
anti-self-dual two-forms.

In the following we will take the normalization
constant $C$ to be equal
to $C=g^{b}$, where $b = \dim H_2(M^4)$,
because with this choice we will find that the partition function is
(almost) invariant under (a maximal subgroup of) the $SL(2,\Z)$
duality group.
To verify that this normalization is correct one could for example
use the relation
\be
g^2{\partial{}\over \partial g^2} \log Z =
\int \langle F\wedge\,{}^*\!F\rangle
\ee
and calculate the right-hand side using an appropriate
regularization procedure.  We have not completed this calculation
but, by analogy with the 2d Gaussian model,
we expect that the regulated one-point function of the
marginal operator $F\,\wedge\!{}^*F$ contains metric-dependent terms
such as the Euler class. It is likely that this leads to the wanted
result
for the normalization constant $C=g^{b}$.

\newsubsection{Duality properties of the partition function}
To derive the behaviour of the full partition function $Z$ under the
group of
duality transformations (\ref{SL2Z}) we only have to consider its
action
on the classical sum $Z_{cl}$ because this is the only part that
depends on
the couplings $g$ and angle $\theta$. By repeatedly using the Poisson
resummation formula
$$
\sum_m f(m)=\sum_n \int\! dx e^{2\pi i n x} f(x)
$$
it is in principle straightforward to compute the action of
the $SL(2,\Z)$ on the sum $Z_{cl}$. The only ingredients that
are used in the calculation are the relation (\ref{QG})
and the fact that $Q_{IJ}$ is unimodular.

It turns out that the partition function $Z$ is in general not
$SL(2,\Z)$-invariant. To describe its transformation properties under
duality,
let us first introduce the generalized partition sum
\be
\label{Zthph}
Z\left[\begin{array}{c}\vec{\theta}\\ \vec{\phi}\end{array}\right]=
g^{-b}
\sum_{m^I} e^{-S[m^I+\theta^I]+ 2\pi i(m^I+\theta^I) Q_{IJ} \phi^J},
\ee
where $S[m]$ is the same quadratic expression given in (\ref{Sm})
and the `characteristics' $\vec{\theta}$ and $\vec{\phi}$ are
half-integers.
The physical interpretation of $\vec{\theta}$ and $\vec{\phi}$ is
that
they represent half-integer shifts in the quantization rule of the
magnetic
and `electric'  fluxes through the homology cycles. Notice that
$Z\left[0\atop 0\right]$ coincides with the Maxwell partition
function,
where we dropped the determinants. We find that under $SL(2,\Z)$
these
partition sums
transform as:
\be
\label{newZ}
Z\left[\begin{array}{c}\vec{\theta}\\ \vec{\phi}\end{array}\right]\ra
\epsilon e^{i\varphi} Z\left[\begin{array}{c}\vec{\theta}'\\
 \vec{\phi}'\end{array}\right]
\ee
where
\be
\label{newtheta}
\left[\begin{array}{c}  \vec{\theta}' \\
\vec{\phi}'\end{array}\right]=
\left(\begin{array}{cc} a & b \\ c & d\end{array}\right)
\left[\begin{array}{c}  \vec{\theta}\\  \vec{\phi} \end{array}\right]
+ \hf
\left[\begin{array}{c}  ab \vec{Q}\\  cd\vec{Q} \end{array}\right],
\ee
$\epsilon$ is some eighth root of unity and
$\varphi=\hf Q^I{}_I\, {\rm arg}(c\tau+d)$. Note that the phase
$\varphi$
depends only
on the topological data contained in $Q_{IJ}$ and is independent of
$G_{IJ}$.
The components of the vector $\vec{Q}$ are given by the diagonal
elements
$Q^{II}$ of the (inverse) intersection form.
The transformation rule (\ref{newZ}) is very similar to the modular
properties of
theta functions associated with 2d Riemann surfaces.
In section 4 it will become clear that this similarity is not
just a coincidence. When the diagonal elements $Q^{II}$ of the
intersection
form are not all even, the Maxwell partition sum $Z\left[0\atop
0\right]$
is not invariant under the $SL(2,Z)$ duality group.
Only on four-manifolds with an
even intersection form does one find that the partition sum is
duality invariant
up to a phase. This fact and the  transformation rules of the
partition sum
can be naturally understood from the quantum properties of the
electric and
magnetic fluxes.

\newsubsection{Flux quantization}
Let us introduce the `electric' and magnetic flux operators
\be
\label{fluxes}
\Phi_m^I = {1\over 2\pi}\int_{\Sigma_I} F, \qquad\qquad
\Phi_e^I = {1\over 2\pi}\int_{\Sigma_I} F_D,
\ee
where
\be
\label{Adual}
F_D\equiv 2\pi i {\delta S\over \delta F}\ =\ {4\pi i\over g^2}{}^*F+
{\theta\over 2\pi} F\qquad{}
\ee
is the dual field strength. Here we have chosen the definition of
$\Phi_e$ so that it transforms nicely under duality, but strictly
speaking it is a linear combination of electric and  magnetic flux.
Under $SL(2,Z)$ the electric and magnetic fluxes transform as
\be
\label{QtoQ}
\left(\begin{array}{c}  \Phi_m \\  \Phi_e\end{array}\right)\ra
\left(\begin{array}{cc} a & b \\ c & d\end{array}\right)
\left(\begin{array}{c}  \Phi_m\\  \Phi_e\end{array}\right)
\ee
By a slight modification of the calculation of section 2.1 it can be
shown
that the expressions (\ref{Zthph})  represent the
(conveniently normalized) expectation values
\be
Z\left[\begin{array}{c}\theta\\ \phi\end{array}\right]=\Bigl \langle
\exp 2\pi i[\phi^IQ_{IJ}\Phi_m^J]\Bigr \rangle_{\theta^I}
\ee
where the subscript $\theta^I$ indicates that the quantization
condition
for the magnetic fluxes is $\Phi_m^J\in \Z\!+\!\theta^I$.
Comparing (\ref{QtoQ}) with the homogeneous term in
the transformation rule (\ref{newtheta}) suggests that $\phi^I$ must
represent a shift in the electric flux quantization. This
interpretation
of $\phi^I$ as well as the mysterious inhomogeneous term in
(\ref{newtheta}) follow from the fact that, as quantum operators,
$\Phi_e$
and $\Phi_m$ do not commute when the corresponding surfaces have a
non-zero intersection. We find
\be
\label{fluxcom}
[\Phi_e^I,\Phi_m^J] ={1\over 2\pi i} Q^{IJ}.
\ee
This can be derived, for example, from the short-distance
properties of the two-point function $\langle F_D F\rangle$ by
imitating
the technique of radial quantization familiar from two-dimensional
conformal
field theory.  The transformation properties of the partition
functions
can now be understood by interpreting (\ref{QtoQ}) as a
canonical transformation in the quantum Hilbert space of the flux
operators.
The result (\ref{newtheta}) precisely describes the
unitary transformation of the eigenstates
$|\vec{\theta},\vec{\phi}\rangle$
of the exponentials $e^{2\pi i\Phi^I_m}$ and $e^{2\pi i \Phi^I_e}$
with
eigenvalues $e^{2\pi i \theta^I}$ and $e^{2\pi i \phi^I}$.
Once this identification is made, it becomes a simple quantum
mechanics
exercise to derive the transformation rule (\ref{newtheta}): it
basically
follows from the CBH formula:
$$
e^{2\pi i (a \Phi^I_e+b\Phi^I_m)} = (-1)^{ab Q^{II}}
\,e^{2\pi i a \Phi^I_e}\,e^{2\pi i b\Phi^I_m}.
$$

\newsection{Relations with 2D Toroidal Models\footnote{The results
of this and the preceding section have been reported at the Strings
'95
conference at USC and at the Spring School in Trieste,
and appear to have some overlap with \cite{Wittenprep}.}}
\vspace{-7mm}
\newsubsection{Self-dual Lorentzian lattices}
The classical partition sum of Maxwell theory on a four-manifold
has a close similarity to that of a two-dimensional toroidal model.
We can make the correspondence almost perfect by noticing that
the partition sum can be rewritten as
\be
\label{latticesum}
Z_{cl}(\tau,\taubar)={\oversea}\sum_{(p^+,p^-) \in \Gamma_{b^+,b^-}}
\exp \Bigl[i\pi \tau (p^+)^2 -i\pi\taubar (p^-)^2\Bigr].
\ee
where  we sum over  a  self-dual lorentzian lattice with `signature'
$(b^+,b^-)$. Here $b^+$ ($b^-$) is the number of (anti-)self dual
harmonic two-forms, and coincides with the number of positive
(negative) eigenvalues of the intersection form $Q_{IJ}$.
To be more precise,  we can represent the lattice
in terms of a set of generators
\be
\Gamma_{b^+,b^-}=
\bigoplus_{I}\,\Z\,
({\bf e^+}_I\oplus {\bf e^-}_I),
\ee
which are related to the $Q_{IJ}$ and $G_{IJ}$ via
\be
\label{ee}
\hf(G_{IJ}\pm Q_{IJ}) =
\sum_{i=1}^{b^\pm} ({\bf e^\pm}_{I})^{i}({\bf e^\pm}_J)^i.
\ee
The expression (\ref{latticesum}) is identical to the
partition sum of a 2d toroidal model used for
string compactifications, but without the powers of the
Dedekind $\eta$-function that represent the oscillator
modes of the string. We should note also that the lattices
that arise for four-manifolds are integral but, unlike those
used for toroidal string compactifications,  not always even.

The classical partition sum is a function of
$b^+\times b^-$ moduli parameters that parametrize the
shape of the lattice and take values on the coset space
$$
{\cal M}_{b^+,b^-}=
SO(b^+)\times SO(b^-)\backslash SO(b^+,b^-)/ O(b^+,b^-,\Z).
$$
A well-known example is ${\cal M}_{19,3}$, which represents
the moduli space of $K^3$-manifolds. The symmetry of the partition
sum
under the discrete group $O(b^+,b^-,\Z)$  ensures
its invariance under  the mapping class group of the four-manifold.
The fixed points of elements of $O(b^+,b^-,\Z)$ correspond to
four-manifolds with accidental discrete symmetries. It is known that
the partition sum at these enhanced symmetry points often contains
purely $\tau$-dependent (or $\taubar$-dependent) lattice sums.
These `characters' represent the contributions of abelian
instantons ($=$ purely self-dual (or anti-self-dual) solutions
of Maxwell's equations).

\newsubsection{Some examples}
As an illustration let us discuss a few simple examples.
The basic example of a manifold with an odd-intersection
form is $CP^2$ for which $b^+=1$ and $b^-=0$. The classical
Maxwell partition sum on this manifold is given by a
Jacobi theta-function
$$
Z_{cl}(\tau)_{\strut CP^2} ={\oversea}\theta_3(\tau)=
{\oversea}\sum_m e^{i\pi\tau m^2}.
$$
This partition-function is invariant only under $\tau\ra \tau+2$,
and up to a phase under $\tau \ra -1/\tau$, provided we choose the
normalization constant to be $1/C=\sqrt{\Im\tau}$.

The simplest non-trivial example of an `even' manifold is
$S^2\times\! S^2$ which has intersection form $Q=
{\mbox{\footnotesize${\left(\begin{array}{cc}0&1\\1&0\end{array}\right)}$}}.$
When we specialize our result to this case we find the
Maxwell partition sum for electro-magnetism on $S^2\times S^2$
is identical to the momentum sum for the $c\!=\!1$ gaussian model
$$
 Z_{cl}(\tau,\taubar, R)_{\strut S^2\times S^2} ={\oversea}
\sum_{m,n} e^{{i\over 2}\pi\tau ({n\over R}+mR)^2}e^{-{i\over
2}\pi\taubar
({n\over R}-mR)^2},
$$
where the `radius' $R$ equals the ratio of the size of the two
spheres.
Note that the familiar  $R\ra {1/ R}$-symmetry
is just a consequence of the invariance under exchange of the
two spheres. Hence, in this context, $R\ra {1/ R}$-duality
corresponds to a kind of `4d modular invariance', or, more
precisely, invariance under the mapping class group  $\Z_2$ of
$S^2\times S^2$.

These observations can be generalized to manifolds with higher
dimensional second co-homology, such as the connected sum of
$N$ copies of the
$S^2\! \times S^2$-manifold.
On this manifold, which we simply denote as $N(S^2\! \times S^2)$,
we can choose a canonical basis of $\Aa$-
and $\Bb$-cycles $\Sigma_i$ and $\widetilde{\Sigma}^j$ such that the
corresponding two-forms $\alpha_i$ and $\beta^j$ have  intersection
\be
\label{intersect}
 \int_\XX\! \alpha_i\wedge\beta^j = \delta_i{}^j
\ee
with $i,j=1,\ldots, N$. All other components of the intersection
form $Q$ vanish. By making use of the relation (\ref{QG}) we find
that
the metric on these two-forms takes the form
\be
\label{GandB}
\int_\XX\! \beta^i\wedge{}^\ast\beta^j = G^{ij},\qquad\qquad
\quad \int_\XX\! \alpha_i\wedge{}^\ast\beta^j = B_i{}^j
\ee
and
\be
\!\!\!\!\!\!\!\!\int_\XX\! \alpha_i\wedge{}^\ast\alpha_j =
G_{ij}-B_{ik} G^{kl} B_{lj},\qquad\quad
\ee
where $G_{ij}$ is symmetric and positive definite  and the matrix
$B_{ij}\equiv B_i{}^k G_{kj}$ is antisymmetric.
The classical action for the saddle-point configurations
may thus be written as
\be
\label{Smn}
S[m,n] =
{8\pi^2\over g^2} \Bigl[m^i G_{ij} m^j +
(n_i - B_{ik} m^k) G^{ij}(n_j - B_{jl} m^l)\Bigr] -i\theta m^i n_i.
\ee
This expression exactly coincides with the spectrum of
vertex operators for a toroidal model  with
constant metric $G_{ij}$ and anti-symmetric tensor field $B_{ij}$.
The properties of the resulting partition sum have
been well studied in this string context (see e.g.
\cite{giveonetal}), and
we gratefully make use of this.

For $N(S^2\times S^2)$, or any other four-manifold with the same
intersection form, we can rewrite the partition in a manifestly
duality-invariant way by performing a Poisson resummation over the
fluxes through the $\Bb$-cycles $n^i$.
This gives
\be
\label{partsum}
Z_{cl}(G,B,\tau,\taubar)=\sum_{m,n} e^{-2\pi E[m,n]}
\ee
where
\be
\label{Emn}
E[m,n] =  {1\over \Im\tau}(n^i+\tau m^i) (G_{ij}+B_{ij})(n^j+\taubar
m^j).
\ee
The integers $n^i$ and $m^i$ now represent the electric and magnetic
flux through the $\Aa$-cycles $\Sigma_i$.
To get to this $SL(2,\Z)$-invariant representation of the
partition sum, we had to choose a canonical decomposition of the
two-cycles.
Of course, we would get the same partition sum if we had relabelled
the basis of two-spheres without changing the intersection form.
This fact gives rise to the familiar $O(N,N,\Z)$ symmetry.

\newsubsection{Correlators through factorization.}
It is also interesting to study the partition function on certain
degenerate four-manifolds.
This should give information on the spectrum of
states of the theory, and the correlation function of observables.
It turns out that the relevant type of degenerations are those in
which
a two-cycle  shrinks to size zero. When an $\Aa$-cycle $\Sigma_i$
is pinched the corresponding metric-element $G_{ii}$ blows up and
goes to infinity. We deduce from (\ref{partsum}) that for $G_{ii}\ra
\infty$
the term $Z[m,n]$ in the partition sum labelled by $m^i$ and $n^i$
is suppressed by an exponential factor $e^{-2\pi G_{ii} \Delta_{mn}}$
with
$$\Delta_{mn} = |n+m\tau|^2/\Im\tau.$$
In analogy with 2d conformal field theory we would like to
interpret $\Delta_{mn}$ as the `scaling dimension' of the
operators in the theory. Indeed, when we `pinch' all `$\Aa$'-cycles
the partition function can be seen to go over in the correlation
function of
the abelian versions of the Wilson-'t Hooft line operators,
\be
\label{factor}
 Z[m, n]\ra \prod_i e^{-2\pi G_{ii}\Delta_{m_in_i}}\Bigl\langle
\prod_i W_{m_in_i}(C_i)\Bigr\rangle.
\ee
The observables $W_{mn}$ are represented by
\be
\label{Wn}
W_{mn}(C)=\exp\, i \Bigl(
n\oint_C A+m\oint_C A_D\Bigr) ,
\ee
where  $A_D$  is the dual gauge field whose field strength $F_D=dA_D$
is given in (\ref{Adual}). The integers $n$ and $m$ are the electric
and magnetic
charge resp. Assuming that the loop $C$ is contractible we can
rewrite the line integrals over $A$ and $A_D$ as a surface integral
of the $F$ and $F_D$ over a disk $D$ with boundary $\del D=C$.
Then, because the functional integral is just a gaussian, we can
formally
express the correlation functions of the observables
$W_{m,n}$ in the two-point function
of the field strength $F$. In this way one finds
\be
\label{correl}
\Bigl\langle \prod_i W_{n_i,m_i}(C_i)\Bigr\rangle = \prod_{i\neq j}
\exp {2\pi\over \Im \tau}(n_i+\tau m_i)\Eij (n_j+\taubar m_j)
\ee
where
\be
L(C_i,C_j)= {g^{-2}}\int_{D_i}\int_{D_j} \langle \!F^+\, F^-\rangle,
\ee
with $F^\pm=\hf (F\pm {}^*F)$
and $\del D_i = C_i$. Notice that this has indeed a form identical to
the
term in the partition function labelled by $m_i$ and $n_i$, as would
be
expected from the factorization equation (\ref{factor}).

\newsection{Duality from Dimensional Reduction}
\vspace{-.6cm}
\newsubsection{Self-dual $2$-form theory in $d=6$ and its
reductions.}
In this section we further clarify the relation between the duality
and
modular symmetries of Maxwell theory and the two-dimensional toroidal
models
by showing that both these theories can be regarded as
dimensional reductions of the same theory, namely
of a six-dimensional theory describing a $2$-form field $C$ with
self-dual field strength $H=dC$: Maxwell
theory is obtained by compactifying the 6d self-dual theory on
a torus, while the toroidal models arise through compactification
on $M^4$.

First let us explain how to describe the
self-dual 2-form theory in $d=6$.
 As a starting point, let us consider the following
first order action\footnote{For definiteness, we restrict our
attention to the
6d theory, but the generalization to self-dual $2p$-forms in $4p+2$
dimensions
should be obvious.}
\be
\label{S0}
S_{6d}={1\over 2\pi i} \int dC\wedge H + {1\over 4\pi} \int H\wedge
{}^* H.
\ee
where $H$ is a three-form field and the field $C$ is a
two-form satisfying the flux quantization condition
$\int_\Xi dC \!\in\! 2\pi\Z$
for all three-cycles $\Xi$.  Integrating out $H$ implies that
$H=i\,{}^*\!dC$
and gives the standard free action for a $C$. Integrating
out $C$ implies  $H=dC_D$ and gives the dual action for the
two-form field $C_D$. In (\ref{S0}) we have chosen the coupling
constant at its self-dual
value so that the action for $C$ and its dual $C_D$ are identical.

To construct an action for the {\it self-dual} 2-form field we now
take this
first order action and identify {\it half} of the components of $H$
with those of $dC$. The only problem with this procedure is that one
has
to give up manifest covariance of the theory. For example, one way
to do this is  by choosing a global vector field $V$
(=`time'-direction)
and to equate all the `spatial' components of $H$ with those of $dC$,
or
in a more invariant notation  \be \label{polar} i_V(H- dC)=0.\ee
This leads to the non-covariant action for self-dual $2$-forms of
Henneaux and
Teitelboim \cite{HT}. The physical content of the theory should not
 depend on the choice of the vector field $V$, and hence there must
be a `canonical transformation' that relates different choices for
$V$.
This is somewhat analogous to the `choice of polarization' in quantum
mechanics,
and suggests a possible reinterpretation of this procedure in terms
of
geometric quantization.

We now describe how the above procedure leads to the Maxwell action
when one reduces to four dimensions by compactifying on a torus.
Let us choose complex coordinates $z$ and $\zbar$ on
the torus $T^2$ with modular parameter $\tau$.  The dimensional
reduction
is performed by taking the following ansatz for the fields $H$ and
$C$
\ba
\label{ansatz1}
H\is{1\over \Im \tau}\Bigl[(F_D-\tau F)dz  + (F_D-\taubar
F)d\zbar\Bigr],\\[3mm]
C\is {1\over \Im \tau}\Bigl[(A_D-\tau A)dz  + (A_D-\taubar
A)d\zbar\Bigr]\nonumber
\ea
where $F$ and $F_D$ are four-dimensional two-forms which for the
moment
are unrelated to the four-dimensional gauge fields $A$ and $A_D$.
Inserting this ansatz into the action (\ref{S0}) and performing the
integrations over $z$ and $\zbar$ gives
\be
\label{S4d}
S_{4d}= {1\over 2\pi i}\int (dA_D\wedge F+ dA\wedge F_D) +
{1\over 4\pi \Im \tau} \int (F_D-F\tau)\wedge{}^*\!(F_D-F\taubar).
\ee
To proceed we now choose our vector field $V$ to be along the
$b$-cycle of the
 torus. The condition  (\ref{polar}) then gives $F=dA$.
The next step is to integrate out the field $F_D$ from the
action(\ref{S4d}).
One easily checks that this gives the Maxwell action (\ref{action}).
Notice that by the field equations
$F_D$ becomes identified with the dual field strength
$F_D = i \Im \tau {}^* F+ \Re \tau F$ introduced in (\ref{Adual}).
And
with this one can verify that the field $H$ in (\ref{ansatz1})
is indeed self-dual.

In this construction the $SL(2,\Z)$-duality symmetry coincides with
the modular group of the internal torus, which is
a remnant of 6d-covariance. The fact that the duality
symmetry is not manifest is because our choice of the
vector field $V$ breaks the modular symmetry. Different
choices for $V$ are related by modular transformations: for example,
if we choose this vector in the direction of the $a$-cycle
we would have obtained the dual action with coupling $-1/\tau$.
An other choice would be to take the vector in a preferred direction
in the 4d space(-time). This would lead to the
duality invariant but non-covariant action of Sen and Schwarz
\cite{SenSchw}.
Our results suggest that it is indeed impossible to have manifest
duality and covariance at the same time: otherwise all
partition functions would have been invariant under duality.
The deviations of duality invariance are thus directly related
to the global gravitational anomalies of the 6 dimensional theory.

Let us now explain how one obtains the toroidal model by reducing
the self-dual 6d theory to $d=2$ on a simply-connected
internal manifold $M^4$. For definiteness we assume that
$M^4$ has an even intersection form and
$b^+=b^-$.
The reduction of the three-form field
strength to 2d is performed by imposing $d_4 {}^*H=0$,
where $d_4$ denotes the exterior derivative on $M^4$.
Thus we can write
\be
H=\sum_i \alpha_i dX^i +\beta^i \Pi_i,\nonu
\ee
where $d$ is now the exterior derivative in the two un-compactified
dimensions and  $\alpha_i$ and $\beta^i$ are the same harmonic
two-forms
that we introduced in section 3. We now als assume that there exists
a
vector field $V$ with $i_V\beta^i=0$. With this choice of $V$ the
condition
(\ref{polar}) implies that the field $X^i$ is identified with the
periods of
the two-form field: $X^i = \int_{\Sigma_i} C$. Inserting this ansatz
for $H$ and
$C$ into the action (\ref{S4d}) and performing the integrations over
$M^4$
gives the two-dimensional action
\be
S_{2d}= {1\over 2\pi i}\int \Pi_i \wedge dX^i +{1\over 4\pi}
\int \Bigl[dX^i G_{ij} {}^*dX^j + (\Pi_i - B_{ik} dX^k)\wedge
G^{ij}\,{}^*\!(\Pi_j - B_{jl}dX^l )\Bigr].
\ee
Self-duality of the field strength
$H$ implies $\Pi_i = i G_{ij}\,{}^* dX^j + B_{ij} dX^j$ which is
indeed one of  the field equations that follow from $S_{2d}$.
Finally, integrating out $\Pi_i$ leads to the action of the 2d
toroidal model
\be
S_{2d}
= {1\over 2\pi i }\int  dX^i\wedge (i G_{ij}\,{}^* dX^j + B_{ij}
dX^j).
\ee
Again  the hidden duality symmetries of the dimensionally reduced
theory
are directly related to the symmetries of the internal
compactification manifold. We further note that if we reduce on
a four-manifold with $b^+\neq b^-$,  we will find a theory
with unequal left- and right-moving bosons: for example for
$K^3$ we get $19$ left-movers and $3$ right-movers.

The fact that the toroidal model and Maxwell theory have the same
classical partition sum can now be understood as follows: the
ans\"atze that we used for $H$ to reduce to $d=4$ and to $d=2$
are compatible with the classical solutions of the full 6d theory.
Thus the classical solutions of Maxwell theory and those of the
toroidal
models can be extended to the same self-dual three forms on
$M^4\times T^2$,
and thus are in one-to-one correspondence with the classical
solutions of the 6d theory.
This makes clear that the classical partition sum for all these three
theories are indeed identical.


\newsubsection{A simple topological theory in $d=7$ and its
dimensional
reductions.}
Consider the following  topological theory in 7 dimensions
\be
\label{Stop7}
S_{7d} = {1\over 2\pi}  \int H \wedge d H
\ee
where $H$ is an (unconstrained) three-form. The action (\ref{Stop7})
is
identical to the level $k=1$ $U(1)$ Chern-Simons theory, except that
we have
replaced the abelian gauge-field with a three-form. It is known that
the
Hilbert space of the $U(1)_{k}$ CS-theory may be identified with the
characters
($=$ chiral partition functions) of the chiral boson at $k$ times
the self-dual radius. In a similar way one can show that the Hilbert
space corresponding to $S_{7d}$ is related to the self-dual
2-form theory in $d=6$.  The only subtle point concerns the gauge
symmetry: in order to achieve the correspondence with the 2-form
theory at its self-dual coupling the abelian gauge symmetry
$H\ra H+dC$ must be `compact': this
means that $C$ is allowed to be multi-valued, as long as it
has integral periods $\int_\Xi dC\in 2\pi\Z$ for
all three-cycles $\Xi$.

Let us now consider this topological theory on a seven-manifold of
the
type $ T^2\times M^4\times \R$, where we interpret $\R$ as time.
Now, in a similar way as in subsection 4.1 we can dimensionally
reduce
the theory in various ways.
First, the reduction to $d=3$ gives a familiar theory: by using the
ansatz $H=\sum \alpha_I A^I$, where $\alpha_I$ are again the
 integral harmonic two-forms on $M^4$  we reduce the
theory to $3d$ abelian Chern-Simons theory on $T^2\times \R$
\be
S_{3d} = {1\over 2\pi} \int A^I \wedge  Q_{IJ} dA^J
\ee
with integral coupling constants $Q_{IJ}$ given by
the intersection form of the four-manifold. It is well known that
the Hilbert space of this theory
is related to the 2d toroidal conformal models  \cite{CS}.
In a completely analogous way it is shown that the
$5$-dimensional topological theory
\be
S_{5d} = {1\over 2\pi} \int F_D \wedge d F,
\ee
which is obtained from $S_{7d}$ by inserting the ansatz
(\ref{ansatz1}),
is related to 4d Maxwell theory. The fields $F$ and $F_D$ in $S_{5d}$
are independent and unconstrained two-forms, but, by the
field equations and after quantization, $F_D$ and $F$ become
identified with the Maxwell field strength and its dual.

A curious fact about these topological dimensional reductions is
that they do not reduce the number of physical degrees of freedom:
all these theories have a finite number of degrees of freedom
living on a `small phase space' parametrized by two sets of angles
$\theta^I\!\in \![0,2\pi]$ and $\phi^I\!\in\! [0,2\pi]$.
In the 5d topological model on $ M^4\times \R$ the classical
solutions
for the fields $F$ and $F_D$ are of the
form $F\!=2\pi\!\sum \theta^I\alpha_I$,  $F_D\!=2\pi\!\sum
\phi^I\alpha_I$.
In the abelian Chern-Simons theory on $T^2\times\R$ these same
variables
$\theta^I$ and $\phi^I$ represent the $U(1)$-holonomies around the
$a$- and $b$-cycles.
So, without losing any physical degrees of freedom we can even reduce
the theories to a simple quantum mechanical model in $d=1$
that contains all the relevant topological information. The action is
\be
S_{1d} = {2\pi} \int  \theta^I Q_{IJ} d\phi^I.
\ee
After quantization these variables satisfy the canonical commutation
relations $[\theta^I,\phi^J]\!=\!{1\over 2\pi i} Q^{IJ}$, which for
is just the algebra (\ref{fluxcom}) of electric and magnetic fluxes.
As we explained in section 2.3, the $SL(2,\Z)$ duality group
correspond to
the canonical linear transformations of $\theta^I$ and $\phi^I$.
Now let $|\vec{\theta},\vec{\phi} \rangle$
be the simultaneous eigenstate of the exponentials $e^{2\pi
i\theta^I}$ and
$e^{2\pi i\phi^I}$ (notice that they commute), and let us define the
state
$|0\rangle$ by
$
(\theta^I\!+\!\tau \phi^I) |0;\tau\rangle \!=0
$
Then we have
\be
\!\!\langle \vec{\theta},\vec{\phi}|0;\tau\rangle =
Z\left[\begin{array}{c}\vec{\theta}\\ \vec{\phi}\end{array}\right]
\ee
This same overlap can be computed in the `big phase space' in the
the various topological field theories. This yields a functional
integral
representation that, depending on which topological theory we
consider, is
identical to the partition function of the corresponding free field
theories
in terms of $H=dC$, $F=dA$ or $A^I=dX^I$. The result should
be of course the same for all these different cases.

\newsubsection{Generalizations and  string-related models.}
An obvious way to generalize our results is to consider higher
dimensional
theories with (self-dual) forms. For example, we can take
 a $2(p\!+\!q)$-form theory in $d=4(p\!+\!q)\!+\!2$ with
self-dual $2(p\!+\!q)\!+\!1$-form field strength $H=dC$, and
dimensionally reduce this theory down to
$d=4q$ on a internal compactification manifold $X$ with
dimension $4p\!+\!2$. By using an ansatz\footnote{For a related
discussion in supersymmetric theories see \cite{SFetal}} of the form
$H=\sum \alpha_{{}_A}F^{{}^A}+\beta^{{}^A} F_{{}_A}^D$
and following the same procedure as
in section 4.1 we find a $2q\!-\!1$-form theory in $d=4q$ with action
\be
S = {1\over 2\pi i}\int \Bigl( F^{{}^A}_+\Omega_{{}_{AB}}F_+^{{}^B}
-F^{{}^A}_-\overline{\Omega}_{{}_{AB}}F^{{}^B}_-\Bigr),
\ee
where $F^{{}^A}$ are $2q$-form field strengths and $\Omega_{{}_{AB}}$
is the
period matrix of the internal manifold $X$.  This theory possesses
a duality symmetry that is inherited from the mapping class group of
$X$,
and  that acts on the matrix $\Omega_{{}_{AB}}$ as a $Sp(2b,\Z)$
fractional
linear transformation where $2b=\dim H_{2p+1}(X)$. The physical
observables
of this theory are labelled by electric and magnetic quantum numbers
$n_{{}_A}$ and $m^{{}^A}$ and have  $Sp(2b,\Z)$-invariant `scaling
dimensions'
$$
\Delta_{m,n} =
(m_{{}_A}+\Omega_{{}_{AC}}n^{{}^C})(\Im^{-1}\Omega)^{{}^{AB}}(m_{{}_B}+
\Omega_{{}_{BC}}n^{{}^C}).
$$

A particularly relevant case for string theory is $p=q=1$. Namely,
the type IIB superstring has in its 10d effective action a
$4$-form field with self-dual $5$-form field strength. When we
compactify this field  down to $d=4$ on a 6d internal manifold of
the form $K^3 \times T^2$ following the outlined procedure we get
\be
S_{4d}=
{1\over g^2} \int_{\strut \XX}\! F^I \wedge G_{IJ} {}^\ast F^J-
i{\theta\over 8\pi^2} \int_{\strut \XX}\! F^I Q_{IJ}\wedge F^J.
\ee
where $G_{IJ}$ and $Q_{IJ}$ are the `metric' and intersection form
of the {\it internal} $K^3$-manifold, and the couplings $g$ and
$\theta$
come from the internal $T^2$. It is interesting to note that in this
case the partition sum of this Abelian gauge theory is
$SL(2,\Z)$-invariant up to a phase on any four-manifold:
for example, the partition sum on $CP^2$ is given by a
sum over the self-dual Lorentzian lattice $\Gamma_{19,3}$.
The deviation of exact duality is related to the global
gravitational anomaly of the self-dual form in d=10.
In the full string theory these anomalies cancel, and thus
we know that the phase that arises in the duality transformations
in the 4d theory must be cancelled by the other fields in the low
energy action.

The same appears to hold for the kind of
4d abelian gauge theories that arise in toroidal compactifications
of the heterotic string from $d=10$ to $d=4$: also in this case
the couplings and the theta-angles of the various $U(1)$-gauge fields
are precisely right to have duality invariance (up to a phase)
of the resulting 4d effective theory\footnote{I thank A. Sen for
pointing this out to me.}. Our analysis suggest that  the
`duality-anomaly'
is related or proportional to the global gravitational anomaly in 6d
or 10d.
It will be interesting to verify this explicitly
and to check that all obstructions (including phase factors) to
exact $s$-duality cancel in the complete (effective) string
theory on all possible (= differentiable four-dimensional
spin-)manifolds.
This is left for future work.

\bigskip

\begin{flushleft}
{\bf Acknowledgements}
\end{flushleft}

I would like to thank L. Alvarez-Gaum\'e, R. Dijkgraaf, C. Kounnas,
W. Lerche,
S. Mukhi, A. Sen and H. Verlinde for helpful discussions and valuable
comments.
This work is supported in part by a Fellowship of the Royal Dutch
Academy
of Science, and by the Alfred P. Sloan Foundation.

\renewcommand{\Large}{\large}

\end{document}